\documentclass[letterpaper,10pt]{article} 

\usepackage{opticameet3} 
\usepackage {soul}
\usepackage{multicol}

\newcommand\authormark[1]{\textsuperscript{#1}}

\usepackage{mathtools, amssymb, physics, multirow, rotating, amsmath, array, caption}
\usepackage{makecell, tabularx}
\usepackage{enumitem}
\usepackage[colorlinks=true,bookmarks=false,citecolor=blue,urlcolor=blue]{hyperref}

\begin{document}

\captionsetup{width=\textwidth}

\title{Latency of multipartite entanglement distribution in a quantum SDN architecture}


\author{A. Gore\authormark{1*} and A. Beghelli\authormark{1}}

\address{\authormark{1}
Optical Networks Group, Department of Electronic \& Electrical Engineering, University College London, London, WC1E 7JE, UK\\}

\email{\authormark{*}anuj.gore.22@ucl.ac.uk} 

\begin{abstract}
We quantify the latency to distribute multipartite entanglement over a quantum SDN architecture. Results show that classical communication can account for more than 80\% of the delay, highlighting the need for efficient control plane design.
\end{abstract}

\section{Introduction}

Quantum networks will enable the distribution of information via entanglement across long distances, supporting applications in enhanced security and quantum computing \cite{Wehner_2018}. Multi-user quantum applications are enabled by multipartite entanglement distribution (MED) protocols. The latency of these MED protocols, i.e. the time taken to distribute a GHZ state, is key as it impacts both, network performance and feasibility.

Previous work on latency has mostly focused on bipartite entanglement distribution protocols \cite{Shchukin_2019, Brand_2020, Khatri_2019, Kao_2024, Pirandola_2017, Pirandola_2019}. Latency of MED protocols has also been investigated, but with some simplifying assumptions that affect the validity of results for wider networks, such as absence of repeaters \cite{Avis_2023}. Other works use the number of time-steps rather than actual time \cite{Avis_2023, B_uml_2020} or assume that failure in a section restarts the whole system, which increases latency unnecessarily \cite{Bugalho_2023}.



In this paper, we present a novel result for the MED protocol latency and show that classical communication strongly dominates in repeater-based quantum networks. We assumed a quantum Software-Defined Network (SDN) architecture where an SDN controller receives multipartite entanglement requests and orchestrates entanglement distribution among user nodes. Analytically validated Monte-Carlo simulation results show that, for an 8x8 grid network, classical communication accounts for more than 80\% of the latency as soon as the link length exceeds 9 km, highlighting the need to design protocols with minimum control plane signalling requirements.

\section{Quantum SDN Architecture Model}\label{sec:model_arch}

\begin{figure}[h]
     \includegraphics[width=\linewidth]{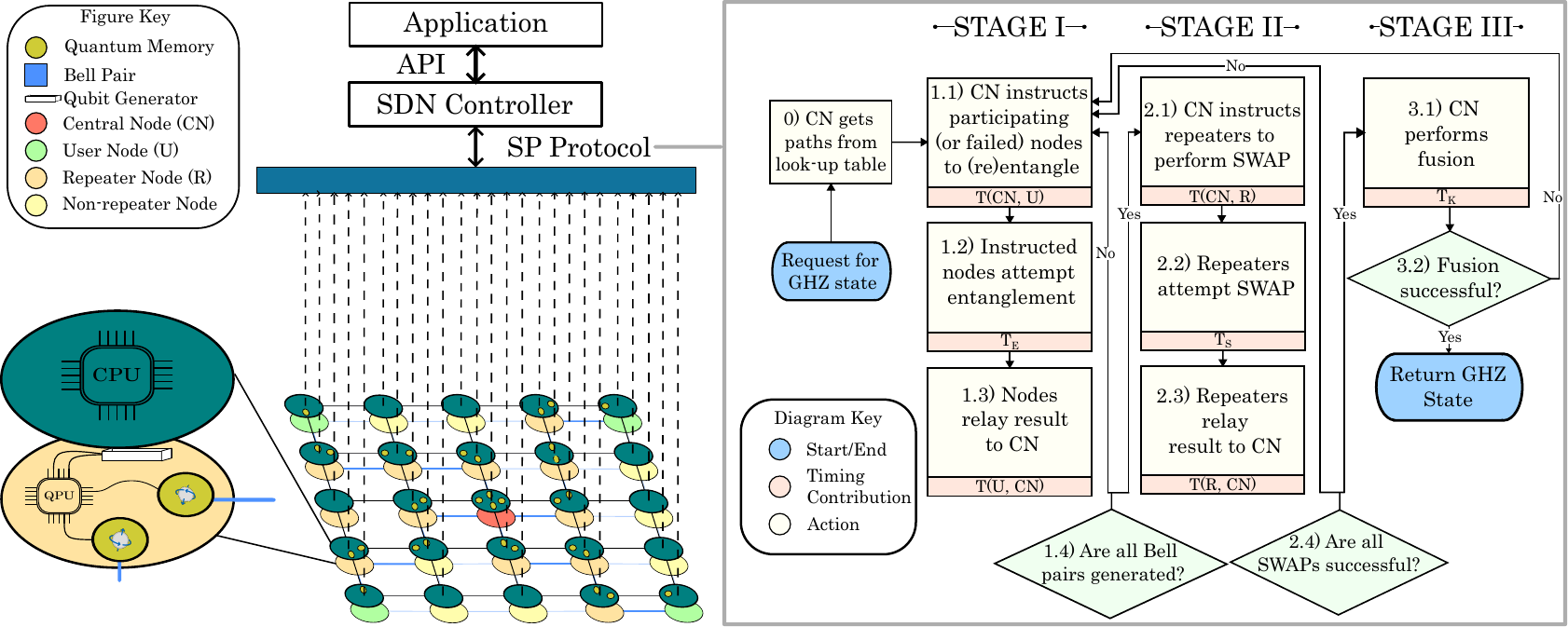}
     \caption[width = \linewidth]{Schematic of a quantum SDN architecture (left) and SP MED protocol (right). $T(A, B)$: signal propagation time from node $A$ to $B$. CN: Central node. R: Repeater node. U: User node. $T_x$: time to execute operation $x$.} \label{diagram:SP_contributors}
\end{figure}

Fig. \ref{diagram:SP_contributors} (left) shows the SDN architecture used to evaluate the MED protocol latency. Network applications send multipartite entanglement requests to the SDN controller via an Application Programming Interface (API), specifying the set of $N$ users needed to be entangled by sharing a $N$-qubit Greenberger-Horne-Zeilinger (GHZ) state. The SDN controller communicates classically with the network nodes (vertically dashed lines) and triggers the execution of the MED protocol. This execution is delegated to the central node (CN), located at the centroid of the users. Once a GHZ state is generated, CN signals the SDN controller, which signals the network application.

In the network, each node is assumed to be made of two layers: a classical electronic layer and a quantum layer. The classical layer is equipped with a classical communication channel to every neighbour node and to its respective quantum layer. It also hosts a copy of the MED protocol. The quantum layer can generate Bell pairs with coherence time (lifetime) $T_c$, perform entanglement swapping (to enable long-distance entanglement) and fusion \cite{nielsen_2010} with success probability $p$, $q$ and $k$, respectively. Heralded Bell pair generation is performed as described in \cite{Krutyanskiy_2023}.

The quantum device is connected to each neighbour node via a physical channel and is equipped with one quantum memory per physical edge to store entanglement. Each physical edge (for quantum and classical communication) is assumed to be a single mode optical fibre, with attenuation being the only source of loss.

\section{Multipartite Entanglement Distribution (MED) Protocol}\label{sec:model_prot}

The execution of the single-path (SP) MED protocol is performed in 3 stages, as shown in Fig. \ref{diagram:SP_contributors} (right). 
    \begin{itemize} [noitemsep,topsep=0pt,parsep=0pt,partopsep=0pt]
        \item[$\ast$] STAGE I: Bell pairs are established between adjacent nodes on pre-calculated routes between CN and users. 
        \item[$\ast$] STAGE II: Entanglement swapping is performed in the repeater nodes to produce long-distance Bell pairs between the CN and the users. Failed swapping requires affected nodes to return to STAGE I. Stages I-II are repeated until all users share a long-distance Bell pair with the CN.
        \item[$\ast$] STAGE III: Fusion operation is performed in the CN. If fusion fails, the protocol returns to STAGE I
    \end{itemize}

\section{Measuring quantum and classical delays: Methodology}\label{sec:method}



\begin{figure}[!ht]
    \centering
    \includegraphics[width=\linewidth]{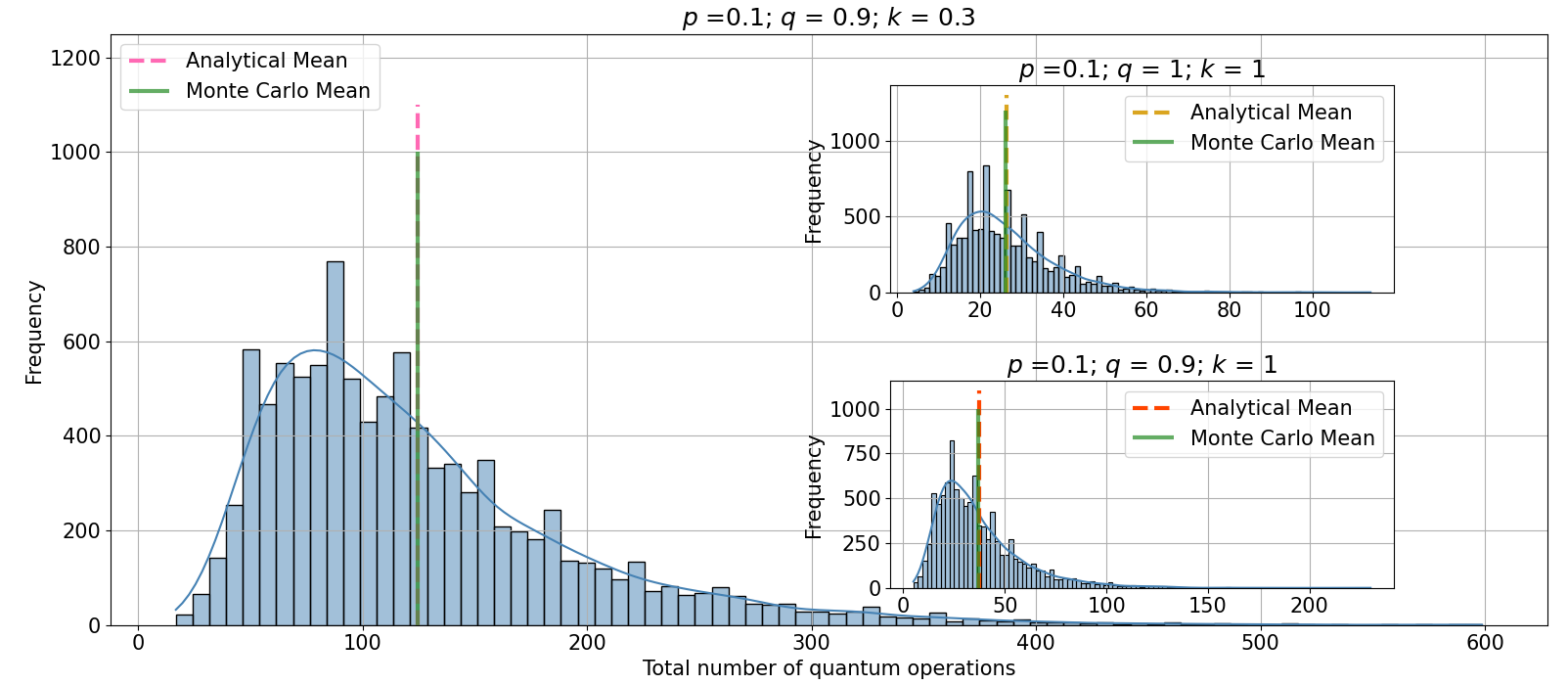}
    \caption{ Simulation histogram and simulation vs. analytical mean values of the number of quantum operations required by the MED protocol to generate a 4-user GHZ state in an $8 \times 8$ grid.}
    \label{fig:mc-validity}
\end{figure}

To account for the probabilistic nature of the quantum operations, we run the MED protocol described in Fig \ref{diagram:SP_contributors} as a Monte-Carlo simulation 10000 times for different values of $p$, $q$ and $k$ and count the total number of times entanglement ($n_{e}$), swapping ($n_{s}$) and fusion ($n_f$) operations are attempted. We verify our Monte-Carlo simulation results with a Markov Chain model, extended from previous work developed for the bipartite case \cite{Brand_2020, Shchukin_2019}. Fig \ref{fig:mc-validity} shows the histogram of the total number of quantum operations required to distribute a GHZ state obtained via the Monte Carlo simulation in an $8 \times 8$ grid topology with the users in the corners and $T_c=\infty$. The vertical lines show the excellent match between the Monte Carlo simulation and Markov Chain model mean values. 

\section{Numerical Results}

The verified Monte-Carlo simulation is then used to obtain latency results for the same grid topology, with distance between adjacent nodes $L \in [1, 29]$ km. We assume a worst-case latency scenario: the users are located at the corner nodes; classical signal latency (signalling for entanglement and swapping) is calculated for the farthest away node/user. We split the total latency between classical ($\tau_{classical}$) and quantum ($\tau_{quantum}$) latency. $\tau_{classical}$ depends on  $n_{e}$, $n_{s}$ and the propagation time of classical signals. Considering the worst case, classical signals travel $2(V-1)\frac{L}{c'}$ for the farthest user, and $2(V-2)\frac{L}{c'}$ for the farthest repeater, with $V$ is the number of vertices traversed from CN to the user, $L$ the length (in km) of one edge and $c'$ the speed of light in fibre. Similarly, $\tau_{quantum}$ depends on $n_{e}$, $n_{s}$, $n_f$ and photon propagation time from Bell pair generation. Entangling, swapping and fusion times are denoted by $\tau_{e}$, $\tau_{s}$ and $\tau_{f}$.
 


Ref. \cite{Krutyanskiy_2023} experimentally demonstrated a single repeater node by entangling a trapped ion and the photon emitted from the excited ion, where swapping is performed via the M\o{}lmer-S\o{}rensen (MS) gate. The Monte Carlo simulation uses the  following values reported in  \cite{Krutyanskiy_2023} : $p$ = $0.018 \times 10^{\frac{-0.2L}{10}}$;$\tau_{e} = 123 \mu s$ and $\tau_{s} = 2157 \mu s$ (without error correction). Ref.  \cite{Kaufmann_2017} generates a GHZ state from Bell pairs on trapped ion qubits by applying three 2-qubit gate rotations, with each gate operation taking $100 \mu s$. Thus, $\tau_{f}=300 \mu s$. To account for experimental imperfections, we assume $q = 0.7$ and $k = 0.5$. Our code is available at \cite{git_repo}.


\begin{figure}[ht]
    \centering
    \includegraphics[width=\linewidth]{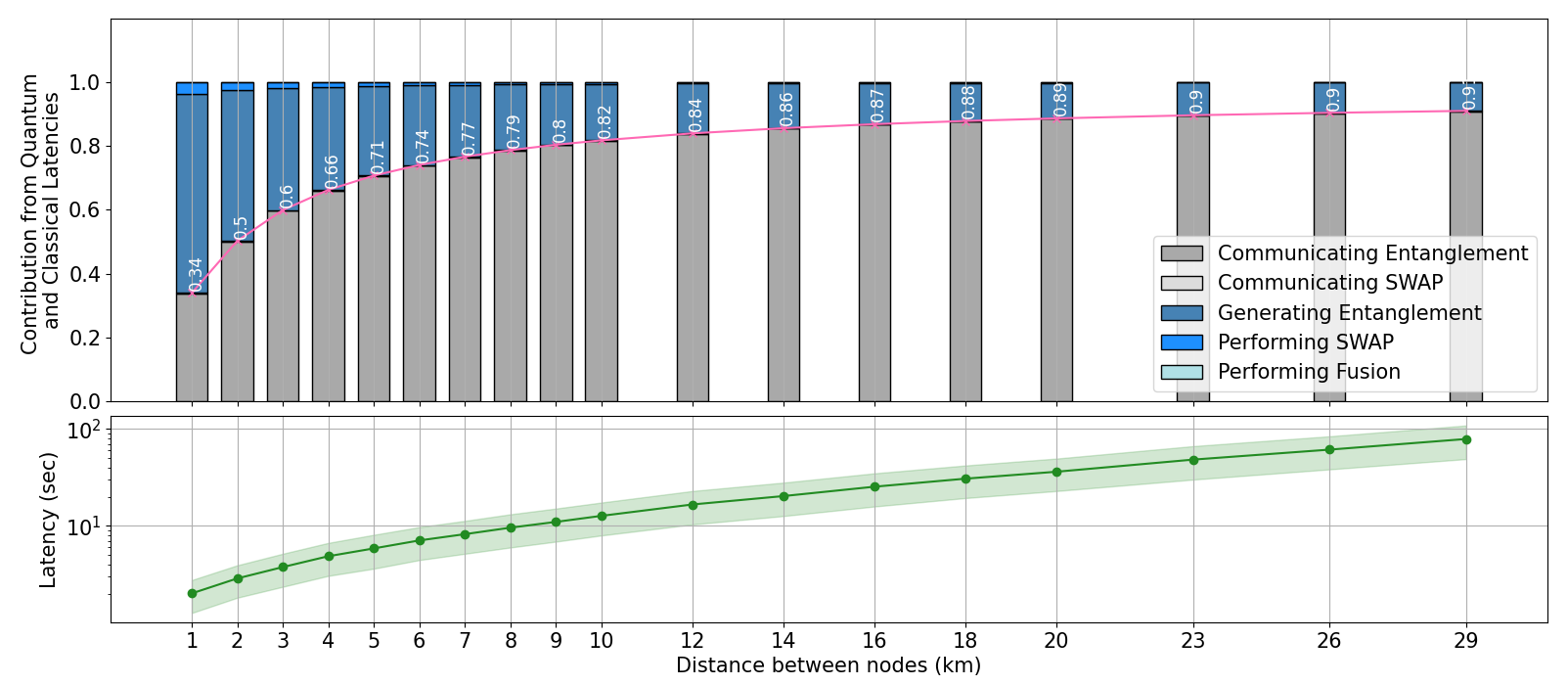}
    \caption{Ratio between quantum (blue) and classical (grey) latency contribution (top plot) and total latency (bottom plot) as a function of link length for GHZ state generation between 4 corner users in an 8x8 grid topology. }
    \label{fig:ratio-cc-qo}
\end{figure}

The top plot of Fig. \ref{fig:ratio-cc-qo}  shows  the contribution of classical and quantum latency. It can be seen that for L $>$ 2 km the majority of the delay is due to classical communication (grey bars) used for entanglement signalling, accounting for  $\geq$ 80\% of latency for L $\geq$ 9 km. While quantum operation related variables depend on experimental details, our analysis (not shown here due to space constraints) shows the same overall trend: classical communication dominates latency. The bottom plot of Fig. \ref{fig:ratio-cc-qo} shows the total latency as a function of link length $L$. It can be seen that latency quickly reaches the second timescales with $L \leq 6$ km. Given current experimental values of $T_c$ for ion-based repeaters ($^{40}$Ca$^+$) below 4 seconds \cite{Chen_2024}, for link lengths $> 4$ km, longer qubit coherence is needed in the studied network. These results highlight the need for protocol design with minimum control plane signalling. \\

\small{\noindent\textbf{Acknowledgements}. Financial support from the NNTI Joint Lab, EPSRC Grant EP/W032643/1 and Innovate UK Quantum Data Centre of the Future (10004793) is gratefully acknowledged.}

\begin{multicols}{2}

\end{multicols}
\end{document}